\documentclass[12pt]{article}
\usepackage{amssymb}

\usepackage{amsmath}
\usepackage{color}
\usepackage[unicode,bookmarks,bookmarksopen,bookmarksopenlevel=2,colorlinks,linkcolor=blue,citecolor=green]{hyperref}

\def\comment#1{}
\input{tcilatex}

\begin{document}

\title{Hydrodynamic chains and a classification of their Poisson brackets}
\author{Maxim V. Pavlov}
\date{}
\maketitle

\begin{abstract}
Necessary and sufficient conditions for an existence of the Poisson brackets
significantly simplify in the Liouville coordinates. The corresponding
equations can be integrated. Thus, a description of local Hamiltonian
structures is a first step in a description of integrable hydrodynamic
chains. The concept of $M$ Poisson bracket is introduced. Several new
Poisson brackets are presented.
\end{abstract}

\tableofcontents

\bigskip

\textit{keywords}: Poisson bracket, Hamiltonian structure, hydrodynamic
chain, Liouville coordinates\textit{.}

\bigskip


\newpage

\section{Introduction}

The theory of the hydrodynamic type systems%
\begin{equation}
u_{t}^{i}=\upsilon _{j}^{i}(\mathbf{u})u_{x}^{j}\text{, \ \ \ }i,j=1,2,...,N
\label{first}
\end{equation}%
integrable by the generalized hodograph method (see \textbf{\cite{Tsar}})
starts from the hydrodynamic type systems equipped with the local
Hamiltonian structure%
\begin{equation*}
u_{t}^{i}=\{u^{i},\mathbf{\bar{h}}\}=[g^{ij}\partial _{x}-g^{is}\Gamma
_{sk}^{j}u_{x}^{k}]\frac{\delta \mathbf{\bar{h}}}{\delta u^{j}},
\end{equation*}%
determined by the Hamiltonian $\mathbf{\bar{h}}=\int h(\mathbf{u})dx$ and
the Dubrovin--Novikov bracket (a differential-geometric Poisson bracket of
the first order, see \textbf{\cite{Dubr+Nov}})%
\begin{equation}
\{u^{i}(x),u^{j}(x^{\prime })\}=[g^{ij}\partial _{x}-g^{is}\Gamma
_{sk}^{j}u_{x}^{k}]\delta (x-x^{\prime })\text{, \ \ \ \ \ }i,j=1,2,...,N,
\label{ha}
\end{equation}%
where the \textbf{flat} metric $g^{ij}(\mathbf{u})$ is symmetric and
non-degenerate, $\Gamma _{sk}^{j}=\frac{1}{2}g^{jm}(\partial
_{s}g_{mk}+\partial _{k}g_{ms}-\partial _{m}g_{sk})$ are the Christoffel
symbols. Such a local Hamiltonian structure can be written via so-called the 
\textit{Liouville} coordinates $A^{k}(\mathbf{u})$, so that the
corresponding Poisson bracket is%
\begin{equation}
\{A^{k}(x),A^{n}(x^{\prime })\}=[\mathcal{W}^{kn}(\mathbf{A})\partial
_{x}+\partial _{x}\mathcal{W}^{nk}(\mathbf{A})]\delta (x-x^{\prime })\text{,
\ \ \ \ \ }k,n=1,2,...,N.  \label{i}
\end{equation}

This paper deals with the hydrodynamic chains (cf. (\textbf{\ref{first}}))%
\begin{equation}
U_{t}^{k}=\underset{n=0}{\overset{k+1}{\sum }}V_{n}^{k}(\mathbf{U})U_{x}^{n}%
\text{, \ \ \ \ \ }k=0,1,2,...,  \label{chain}
\end{equation}%
where all components $V_{n}^{k}(\mathbf{U})$ are functions of the field
variables $U^{m}$ and the index $m$ runs from $0$ up to $k+1$ (except for
possibly \textit{several first} equations, see the end of the next section).
This is a very important class of hydrodynamic chains. Plenty known
hydrodynamic chains (see \textbf{\cite{Benney}}, \textbf{\cite{Blaszak}}, 
\textbf{\cite{Fer+Dav}}, \textbf{\cite{Manas}}, \textbf{\cite{Maks+Egor}}, 
\textbf{\cite{Maks+Kuper}}, \textbf{\cite{Teshuk}}) belong to this class.
So, the hydrodynamic chains (\textbf{\ref{chain}}) are an extension of
hydrodynamic type systems (\textbf{\ref{first}}) to an infinite component
case. Also, these hydrodynamic chains allow the invertible transformations%
\begin{equation}
\tilde{U}^{0}=\tilde{U}^{0}(U^{0})\text{, \ \ }\tilde{U}^{1}=\tilde{U}%
^{1}(U^{0},U^{1})\text{, \ \ }\tilde{U}^{2}=\tilde{U}^{2}(U^{0},U^{1},U^{2})%
\text{, ...}  \label{inv}
\end{equation}

The Poisson brackets (see (\textbf{\ref{ha}}))%
\begin{equation}
\{U^{i},U^{j}\}=[G^{ij}(\mathbf{U})\partial _{x}+\Gamma _{k}^{ij}(\mathbf{U}%
)U_{x}^{k}]\delta (x-x^{\prime })\text{, \ \ \ \ \ \ }i,j,k=1,2,3,...
\label{has}
\end{equation}%
for an infinite component case were considered by I. Dorfman in \textbf{\cite%
{Dorfman}} ($N$ component case was completely investigated by B.A. Dubrovin
and S.P. Novikov in \textbf{\cite{Dubr+Nov}} for the non-degenerate matrix $%
G^{ij}$; its degenerate case was considered by N. Grinberg in \textbf{\cite%
{Grin}}).The Jacobi identity yields the set of restrictions (see \textbf{%
\cite{Dorfman}}, \textbf{Theorem 5.14})%
\begin{eqnarray*}
G^{ij} &=&G^{ji}\text{, \ \ \ \ \ \ \ \ \ \ \ }\partial _{k}G^{ij}=\Gamma
_{k}^{ij}+\Gamma _{k}^{ji}\text{, \ \ \ \ \ \ \ \ \ \ \ \ \ }G^{ik}\Gamma
_{k}^{jn}=G^{jk}\Gamma _{k}^{in}, \\
&& \\
0 &=&\Gamma _{n}^{ij}\Gamma _{k}^{nm}-\Gamma _{n}^{im}\Gamma
_{k}^{nj}+G^{in}(\partial _{n}\Gamma _{k}^{mj}-\partial _{k}\Gamma
_{n}^{mj}), \\
&& \\
0 &=&(\partial _{n}\Gamma _{k}^{ij}-\partial _{k}\Gamma _{n}^{ij})\Gamma
_{p}^{nm}+(\partial _{n}\Gamma _{k}^{mi}-\partial _{k}\Gamma
_{n}^{mi})\Gamma _{p}^{nj}+(\partial _{n}\Gamma _{k}^{jm}-\partial
_{k}\Gamma _{n}^{jm})\Gamma _{p}^{ni} \\
&& \\
&&+(\partial _{n}\Gamma _{p}^{ij}-\partial _{p}\Gamma _{n}^{ij})\Gamma
_{k}^{nm}+(\partial _{n}\Gamma _{p}^{mi}-\partial _{p}\Gamma
_{n}^{mi})\Gamma _{k}^{nj}+(\partial _{n}\Gamma _{p}^{jm}-\partial
_{p}\Gamma _{n}^{jm})\Gamma _{k}^{ni},
\end{eqnarray*}%
which in the Liouville coordinates $A^{i}=A^{i}(\mathbf{U})$ (see (\textbf{%
\ref{i}}) and (\textbf{\ref{inv}}); also \textbf{\cite{Dubr+Nov}}, \textbf{%
\cite{Malt+Nov}}) simplify to%
\begin{eqnarray}
(\mathcal{W}^{ik}+\mathcal{W}^{ki})\partial _{k}\mathcal{W}^{nj} &=&(%
\mathcal{W}^{jk}+\mathcal{W}^{kj})\partial _{k}\mathcal{W}^{ni},  \notag \\
&&  \label{jac} \\
\partial _{n}\mathcal{W}^{ij}\partial _{m}\mathcal{W}^{kn} &=&\partial _{n}%
\mathcal{W}^{kj}\partial _{m}\mathcal{W}^{in}.  \notag
\end{eqnarray}%
In comparison with a finite-component case (hydrodynamic type systems (%
\textbf{\ref{first}})), where elements $\mathcal{W}^{ik}$ depend on \textit{%
all} field variables $u^{n}$ ($n=1,2,...,N$), the elements $\mathcal{W}^{ik}$
in an infinite component case (hydrodynamic chains) depend on a \textit{%
finite} number of moments $A^{k}$ only. Thus, the Liouville coordinates $%
A^{k}$ in the infinite component case is a natural generalization of the
flat coordinates in the finite component case (see \textbf{\cite{Dubr+Nov}}).

\textit{The \textbf{main} problem in a classification of integrable
hydrodynamic chains is the description of the Poisson brackets }(cf. (%
\textbf{\ref{i}})) 
\begin{equation*}
\{A^{k}(x),A^{n}(x^{\prime })\}=[\mathcal{W}^{kn}(\mathbf{A})\partial
_{x}+\partial _{x}\mathcal{W}^{nk}(\mathbf{A})]\delta (x-x^{\prime })\text{,
\ \ \ \ \ }k,n=1,2,...
\end{equation*}

\textbf{Definition}: \textit{The hydrodynamic chain }(\textbf{\ref{chain}}) 
\textit{is said to be Hamiltonian, if it can be written in the form}%
\begin{equation}
A_{t}^{i}=[\mathcal{W}^{ij}(\mathbf{A})\partial _{x}+\partial _{x}\mathcal{W}%
^{ji}(\mathbf{A})]\frac{\delta \mathbf{\bar{H}}}{\delta A^{j}}\text{, \ \ \
\ \ }i,j=0,1,2,...,  \label{liu}
\end{equation}%
where the coefficients $\mathcal{W}^{ij}(\mathbf{A})$\textit{\ satisfy the
Jacobi identity }(\textbf{\ref{jac}}).

The Hamiltonian $\mathbf{\bar{H}=}\int \mathbf{H}dx$ depends on several
first moments $A^{k}$.

\textbf{Remark}: The above local Hamiltonian structures (\textbf{\ref{liu}}%
)\ and corresponding Poisson brackets still are not investigated properly.
However, a lot of publications are devoted to their particular cases (see,
for instance, \textbf{\cite{Dorfman}}, \textbf{\cite{Kuper}}, \textbf{\cite%
{KM}}), some of them will be described below. In this paper I \textbf{%
present the program} of an investigation of these Poisson brackets.

In this paper we consider the very important case%
\begin{equation*}
\mathcal{W}^{kn}=\mathcal{W}_{(M)}^{kn}(A^{0},A^{1},...,A^{k+n-M}).
\end{equation*}%
Corresponding Poisson brackets we call $M-$\textit{brackets}. The simplest
sub-cases $\mathcal{W}^{kn}=\mathcal{W}_{(0)}^{kn}(B^{0},B^{1},...,B^{k+n})$
and $\mathcal{W}_{(1)}^{kn}=\mathcal{W}^{kn}(A^{0},A^{1},...,A^{k+n-1})$ are
considered in details below. The first example of such Poisson bracket%
\begin{equation}
\{A^{k},A^{n}\}=(kA^{k+n-1}\partial _{x}+n\partial _{x}A^{k+n-1})\delta
(x-x^{\prime })  \label{kmb}
\end{equation}%
was found in \textbf{\cite{KM}}, the second example is (see \textbf{\cite%
{Kuper}})%
\begin{equation}
\{B^{k},B^{n}\}=[(k+\beta )B^{k+n}\partial _{x}+(n+\beta )\partial
_{x}B^{k+n}]\delta (x-x^{\prime }).  \label{kb}
\end{equation}%
Infinitely many local Hamiltonian structures for the Benney hydrodynamic
chain (see \textbf{\cite{Maks+wdvv}}) and for the Kupershmidt hydrodynamic
chains (see \textbf{\cite{Maks+Kuper}}) are examples of these $M-$brackets.
The third example%
\begin{eqnarray*}
\{A^{k},A^{n}\}_{2} &=&[(k+1)A^{k+n}\partial _{x}+(n+1)\partial
_{x}A^{k+n}+n(k+1)A^{k-1}\partial _{x}A^{n-1} \\
&& \\
&&+\underset{m=0}{\overset{n-1}{\sum }}(mA^{n-1-m}\partial
_{x}A^{k-1+m}+(k-n+m)A^{k-1+m}\partial _{x}A^{n-1-m})]\delta (x-x^{\prime })
\end{eqnarray*}%
is the second local Hamiltonian structure of the Benney hydrodynamic chain
(see \textbf{\cite{Kuper}}), where the moments $A^{k}$ are no longer the
Liouville coordinates. However, this bi-Hamiltonian structure completely
determines the Benney hydrodynamic chain together with its commuting flows.

The \textbf{main observation} successfully utilized in this paper is that
the Jacobi identity (\textbf{\ref{jac}})%
\begin{eqnarray}
\underset{k=0}{\overset{n+j-M}{\sum }}(\mathcal{W}_{(M)}^{ik}+\mathcal{W}%
_{(M)}^{ki})\partial _{k}\mathcal{W}_{(M)}^{nj} &=&\underset{k=0}{\overset{%
n+i-M}{\sum }}(\mathcal{W}_{(M)}^{jk}+\mathcal{W}_{(M)}^{kj})\partial _{k}%
\mathcal{W}_{(M)}^{ni},  \notag \\
&&  \label{full} \\
\underset{n=0}{\overset{i+j-M}{\sum }}\partial _{n}\mathcal{W}%
_{(M)}^{ij}\partial _{m}\mathcal{W}_{(M)}^{kn} &=&\underset{n=0}{\overset{%
k+j-M}{\sum }}\partial _{n}\mathcal{W}_{(M)}^{kj}\partial _{m}\mathcal{W}%
_{(M)}^{in}.  \notag
\end{eqnarray}%
is a system of nonlinear PDE's which can be solved completely, because all
coefficients $\mathcal{W}_{(M)}^{ij}$ depend on different number of the
moments $A^{k}$ (cf. a similar problem in \textbf{\cite{Maks+Egor}}).
Corresponding illustrative examples are given in this paper for $M=0$ and $%
M=1$.

\textbf{Remark}: Of course, the $M-$bracket is just one among many other
examples. For instance, another new Poisson bracket is%
\begin{equation*}
\{A^{k},A^{n}\}=(A^{k\cdot n}\partial _{x}+\partial _{x}A^{k\cdot n})\delta
(x-x^{\prime }).
\end{equation*}%
All other Poisson brackets (\textbf{\ref{liu}}) and corresponding integrable
hydrodynamic chains will be discussed elsewhere.

At the end of this paper local Poisson brackets are generalized on the
simplest \textit{nonlocal} case (in $N$ component non-degenerate case known
as the nonlocal Hamiltonian structure associated with a metric of constant
curvature \textbf{\cite{Fer+Mokh}}). As example, a one parametric family of
such nonlocal Poisson brackets is presented.

\section{The general case}

The Poisson brackets determining the hydrodynamic chains (\textbf{\ref{chain}%
}) written in the Liouville coordinates%
\begin{equation}
\{A^{k},A^{n}\}=[\mathcal{W}_{(M)}^{kn}(A^{0},A^{1},...,A^{k+n-M})\partial
_{x}+\partial _{x}\mathcal{W}_{(M)}^{nk}(A^{0},A^{1},...,A^{k+n-M})]\delta
(x-x^{\prime })  \label{!}
\end{equation}%
satisfy the Jacobi identity (\textbf{\ref{full}}).

The corresponding hydrodynamic chain (\textbf{\ref{chain}}) is%
\begin{equation*}
A_{t}^{k}=\{A^{k},\mathbf{H}_{M+1}\}_{M}=[\mathcal{W}%
_{(M)}^{kn}(A^{0},A^{1},...,A^{k+n-M})\partial _{x}+\partial _{x}\mathcal{W}%
_{(M)}^{nk}(A^{0},A^{1},...,A^{k+n-M})]\frac{\delta \mathbf{\bar{H}}_{M+1}}{%
\delta A^{n}},
\end{equation*}%
where the Hamiltonian is $\mathbf{\bar{H}}_{M+1}\mathbf{=}\int \mathbf{H}%
_{M+1}(A^{0},A^{1},A^{2},...,A^{M+1})dx$. The Hamiltonian structure is
determined by the Hamiltonian, the momentum and the Casimirs. If a
hydrodynamic chain has one more conservation law density, then this
hydrodynamic chain is integrable, because every extra conservation law
density can be used as an extra Hamiltonian density determining commuting
flow (see \textbf{\cite{Kuper}}). Thus, we have several interesting
sub-cases:

\textbf{1}. $M=1$. The famous example of such a Poisson bracket is the
Kupershmidt--Manin bracket (\textbf{\ref{kmb}}). This Poisson bracket has
the momentum $A^{1}$ and the Casimir (annihilator) $A^{0}$.

If $M\geqslant 1$, then corresponding Poisson brackets have $M$ Casimirs.
Since the Hamiltonian is $\mathbf{\bar{H}}_{M+1}\mathbf{=}\int \mathbf{H}%
_{M+1}(A^{0},A^{1},A^{2},...,A^{M+1})dx$, then the momentum can be chosen as 
$\mathbf{\bar{H}}_{M}\mathbf{=}\int A^{M}dx$. The Casimirs can be chosen as $%
\mathbf{\bar{H}}_{k}\mathbf{=}\int A^{k}dx$ ($k=0,1,2,...,M-1$). Then the 
\textit{auxiliary} (natural) restrictions (``normalization'') are%
\begin{eqnarray}
A^{k} &=&\mathcal{W}_{(M)}^{Mk}(A^{0},A^{1},...,A^{k})\text{, \ \ \ \ \ \ }%
k=0,1,2,...,  \notag \\
&&  \notag \\
0 &=&\mathcal{W}_{(M)}^{sk}(A^{0},A^{1},...,A^{k+s-M})\text{, \ \ \ \ \ }%
0\leqslant s<M\text{, \ \ \ \ \ \ }k\geqslant M-s.  \notag \\
&&  \notag \\
\mathcal{W}_{(M)}^{kn} &=&\mathcal{\bar{W}}_{(M)}^{kn}=\limfunc{const}\text{%
, \ \ \ \ \ \ }k=0,1,2,...,M-1\text{, \ \ \ \ \ }0\leqslant n\leqslant M-1-k.
\label{2}
\end{eqnarray}

Thus, such a hydrodynamic chain has at least $M+2$ conservation laws (for an
arbitrary Hamiltonian density $\mathbf{H}_{M+1}$), where the first $N$
conservation laws of the Casimirs are%
\begin{equation*}
A_{t}^{k}=\partial _{x}\left( \underset{n=0}{\overset{M-k-1}{\sum }}(%
\mathcal{\bar{W}}_{(M)}^{kn}+\mathcal{\bar{W}}_{(M)}^{nk})\frac{\partial 
\mathbf{H}_{M+1}}{\partial A^{n}}+\underset{n=M-k}{\overset{M+1}{\sum }}%
\mathcal{W}_{(M)}^{nk}\frac{\partial \mathbf{H}_{M+1}}{\partial A^{n}}%
\right) \text{, \ \ \ \ }k=0,1,2,...,M-1.
\end{equation*}%
The conservation law of the momentum is%
\begin{equation*}
A_{t}^{M}=\partial _{x}\left( \underset{n=0}{\overset{M+1}{\sum }}(\mathcal{W%
}_{(M)}^{nM}+A^{n})\frac{\partial \mathbf{H}_{M+1}}{\partial A^{n}}-\mathbf{H%
}_{M+1}\right) .
\end{equation*}%
The conservation law of the energy is%
\begin{equation*}
\partial _{t}\mathbf{H}_{M+1}=\partial _{x}\left[ \underset{k=0}{\overset{M+1%
}{\sum }}\underset{n=0}{\overset{M+1}{\sum }}\mathcal{W}_{(M)}^{kn}\frac{%
\partial \mathbf{H}_{M+1}}{\partial A^{k}}\frac{\partial \mathbf{H}_{M+1}}{%
\partial A^{n}}\right] .
\end{equation*}

\textbf{2}. $M=0$. The important example of such a Poisson bracket is the
Kupershmidt brackets (\textbf{\ref{kb}}). These Poisson brackets have the
momentum $A^{0}$ only.

\textbf{3}. $M=-1$. These brackets do not have a momentum and annihilators.

\textbf{4}. If $M<-1$, then an investigation of the Poisson brackets and
corresponding hydrodynamic chains should start from commuting flows
determined by a local Hamiltonian density. For instance, if $M=-2$, then the
first nontrivial Hamiltonian hydrodynamic chain (cf. (\textbf{\ref{chain}}))%
\begin{equation*}
A_{t}^{k}=\underset{n=0}{\overset{k+2}{\sum }}V_{n}^{k}(\mathbf{A})A_{x}^{n}%
\text{, \ \ \ \ \ }k=0,1,2,...
\end{equation*}%
is determined by the first local Hamiltonian $\mathbf{\bar{H}}_{0}\mathbf{=}%
\int \mathbf{H}_{0}(A^{0})dx$.

\textbf{Examples}: The Kupershmidt hydrodynamic chains (see \textbf{\cite%
{Kuper}}) have an infinite set of local Hamiltonian structures determined by 
$M-$brackets (\textbf{\ref{!}}), where the first of them are (for the
indexes $M=0,1,2$, respectively; see \textbf{\cite{Maks+Kuper}})%
\begin{eqnarray*}
\{C^{k},C^{n}\} &=&[(\beta k+1)C^{k+n}\partial _{x}+(\beta n+1)\partial
_{x}C^{k+n}]\delta (x-x^{\prime }), \\
&&
\end{eqnarray*}%
\begin{eqnarray*}
\{B^{0},B^{0}\} &=&\beta \delta ^{\prime }(x-x^{\prime })\text{, \ \ \ \ }%
\{B^{k},B^{n}\}=[(\beta k+1-\beta )B^{k+n-1}\partial _{x}+(\beta n+1-\beta
)\partial _{x}B^{k+n-1}]\delta (x-x^{\prime }), \\
&&
\end{eqnarray*}%
\begin{eqnarray*}
\{A^{0},A^{1}\} &=&\{A^{1},A^{0}\}=\beta \delta ^{\prime }(x-x^{\prime })%
\text{, \ \ \ \ \ \ }\{A^{1},A^{1}\}=(\beta -1)(A^{0}\partial _{x}+\partial
_{x}A^{0})\delta (x-x^{\prime })\text{,} \\
&& \\
\{A^{k},A^{n}\} &=&[(\beta k+1-2\beta )A^{k+n-2}\partial _{x}+(\beta
n+1-2\beta )\partial _{x}A^{k+n-2}]\delta (x-x^{\prime }).
\end{eqnarray*}%
If the first Poisson bracket is the well-known Kupershmidt bracket ((\textbf{%
\ref{kb}}), $M=0$), two others ($M=1$ and $M=2$) are new.

The most interesting class of these Poisson brackets is provided by the
brackets, whose coefficients $\mathcal{W}_{(M)}^{kn}$\ are \textit{%
polynomials} with respect to the moments $A^{k}$. The simplest case is given
by the \textit{linear} Poisson brackets determined by%
\begin{equation*}
\mathcal{W}_{(M)}^{kn}(A^{0},A^{1},...,A^{k+n-M})=e_{(M)}^{kn}A^{k+n-M},
\end{equation*}%
where $e_{(M)}^{kn}$ are some constants satisfying the algebraic system (see
(\textbf{\ref{jac}}))%
\begin{equation*}
(e_{(M)}^{p,k+n-M}+e_{(M)}^{k+n-M,p})e_{(M)}^{nk}=(e_{(M)}^{k,n+p-M}+e_{(M)}^{n+p-M,k})e_{(M)}^{np}%
\text{, \ \ \ \ \ \ }%
e_{(M)}^{sp}e_{(M)}^{k,s+p-M}=e_{(M)}^{kp}e_{(M)}^{s,k+p-M},
\end{equation*}%
possibly connected with infinite dimensional analogue of the Frobenius
algebras (cf. \textbf{\cite{Balin}}). Let us look for particular solution in
the form%
\begin{equation*}
e_{(M)}^{kn}=(Ak+Bn+C),
\end{equation*}%
where $A$ and $B$ are some constants. The substitution of this ansatz in the
above algebraic system yields the Kupershmidt brackets (see \textbf{\cite%
{Kuper}}) determined by an arbitrary value $A$, but $B=0$. More general
linear Poisson brackets%
\begin{equation*}
\mathcal{W}_{(M)}^{kn}(A^{0},A^{1},...,A^{k+n-M})=\underset{s=0}{\overset{%
k+n-M}{\sum }}e_{(M)s}^{kn}A^{s},
\end{equation*}%
were considered by I. Dorfman in \textbf{\cite{Dorfman}}. For instance,%
\begin{equation*}
\{A^{0},A^{0}\}=\varepsilon \delta ^{\prime }(x-x^{\prime })\text{, \ \ \ \
\ \ }\{A^{k},A^{n}\}=\underset{m=0}{\overset{M}{\sum }}\gamma
_{m}[kA^{m+k+n-1}\partial _{x}+n\partial _{x}A^{m+k+n-1}]\delta (x-x^{\prime
}),
\end{equation*}%
where $\varepsilon $ and $\gamma _{k}$ are arbitrary constants.

Higher order homogeneous polynomials create more complicated Poisson
brackets, which can be described by algebraic tools. For instance, the
second nontrivial case is the \textit{quadratic} Poisson brackets determined
by%
\begin{equation*}
\mathcal{W}_{(M)}^{kn}(A^{0},A^{1},...,A^{k+n-M})=\frac{1}{2}\underset{m=0}{%
\overset{k+n-M}{\sum }}e_{(M)m}^{kn}A^{m}A^{k+n-m-M},
\end{equation*}%
where $e_{(M)m}^{kn}\equiv e_{(M)k+n-m-M}^{kn}$ are some constants, which
can be found by direct substitution in the Jacobi identity (see (\textbf{\ref%
{full}})). A corresponding system of algebraic relations is very complicated
and could be investigated in details elsewhere.

\textbf{Remark}: For every Poisson bracket with respect to highest moment $%
A^{M+1}$ the Hamiltonian density $\mathbf{H}_{M+1}$ can be \textit{linear} $%
\mathbf{H}_{M+1}=A^{M+1}+F(A^{0},...,A^{M})$, \textit{quasilinear} $\mathbf{H%
}_{M+1}=G(A^{0},...,A^{M})A^{M+1}+F(A^{0},...,A^{M})$ and \textit{fully
nonlinear}. If $M>0$ just in the \textit{linear} case $\mathbf{H}%
_{M+1}=A^{M+1}+F(A^{0},A^{1})$ the hydrodynamic chain has the form (\textbf{%
\ref{chain}}). In general case the corresponding hydrodynamic chain is%
\begin{equation}
A_{t}^{k}=\underset{n=0}{\overset{M+1}{\sum }}V_{n}^{k}(\mathbf{A})A_{x}^{n}%
\text{, \ \ }k=0,1,...,M-1\text{, \ \ \ }A_{t}^{k}=\underset{n=0}{\overset{%
k+1}{\sum }}V_{n}^{k}(\mathbf{A})A_{x}^{n}\text{, \ \ }k=M,M+1,...  \label{j}
\end{equation}

\section{$M=0$}

We omit the sub-index in all components $\mathcal{W}_{(0)}^{nk}$ of the
Poisson bracket (see (\textbf{\ref{!}}), $M=0$)%
\begin{equation*}
\{B^{k},B^{n}\}=[\mathcal{W}^{kn}(B^{0},B^{1},...,B^{k+n})\partial
_{x}+\partial _{x}\mathcal{W}^{nk}(B^{0},B^{1},...,B^{k+n})]\delta
(x-x^{\prime }).
\end{equation*}%
The Jacobi identity (\textbf{\ref{full}})%
\begin{eqnarray*}
\underset{m=0}{\overset{n+k}{\sum }}(\mathcal{W}^{pm}+\mathcal{W}%
^{mp})\partial _{m}\mathcal{W}^{nk} &=&\underset{m=0}{\overset{n+p}{\sum }}(%
\mathcal{W}^{km}+\mathcal{W}^{mk})\partial _{m}\mathcal{W}^{np}, \\
&& \\
\underset{m=0}{\overset{s+p}{\sum }}\partial _{m}\mathcal{W}^{sp}\partial
_{n}\mathcal{W}^{km} &=&\underset{m=0}{\overset{p+k}{\sum }}\partial _{m}%
\mathcal{W}^{kp}\partial _{n}\mathcal{W}^{sm}.
\end{eqnarray*}%
is a nonlinear PDE system describing a family of the local Poisson brackets
connected with the Hamiltonian hydrodynamic chains.

The existence of commuting hydrodynamic chains%
\begin{eqnarray}
B_{t}^{k} &=&[\mathcal{W}^{kn}(B^{0},B^{1},...,B^{k+n})\partial
_{x}+\partial _{x}\mathcal{W}^{nk}(B^{0},B^{1},...,B^{k+n})]\frac{\delta 
\mathbf{\bar{H}}_{1}}{\delta B^{n}},  \label{mo} \\
&&  \notag \\
B_{y}^{k} &=&[\mathcal{W}^{kn}(B^{0},B^{1},...,B^{k+n})\partial
_{x}+\partial _{x}\mathcal{W}^{nk}(B^{0},B^{1},...,B^{k+n})]\frac{\delta 
\mathbf{\bar{H}}_{2}}{\delta B^{n}},  \label{no}
\end{eqnarray}%
where the Hamiltonians are $\mathbf{\bar{H}}_{1}\mathbf{=}\int \mathbf{H}%
_{1}(B^{0},B^{1})dx$ and $\mathbf{\bar{H}}_{2}\mathbf{=}\int \mathbf{H}%
_{2}(B^{0},B^{1},B^{2})dx$, implies the existence of a hierarchy of
integrable hydrodynamic chains (see (\textbf{\ref{chain}})).

The auxiliary restrictions (``normalization'')%
\begin{equation*}
\mathcal{W}^{0k}\equiv B^{k}
\end{equation*}%
can be obtained by virtue the existence of a conservation law of the
momentum $\mathbf{\bar{H}}_{0}=\int B^{0}dx$. Thus, first two conservation
laws are%
\begin{eqnarray*}
B_{t}^{0} &=&\partial _{x}\left( 2B^{0}\frac{\partial \mathbf{H}_{1}}{%
\partial B^{0}}+(\mathcal{W}^{10}+B^{1})\frac{\partial \mathbf{H}_{1}}{%
\partial B^{1}}-\mathbf{H}_{1}\right) , \\
&& \\
\partial _{t}\mathbf{H}_{1} &=&\partial _{x}\left[ B^{0}(\frac{\partial 
\mathbf{H}_{1}}{\partial B^{0}})^{2}+(\mathcal{W}^{10}+B^{1})\frac{\partial 
\mathbf{H}_{1}}{\partial B^{0}}\frac{\partial \mathbf{H}_{1}}{\partial B^{1}}%
+\mathcal{W}^{11}(\frac{\partial \mathbf{H}_{1}}{\partial B^{0}})^{2}\right]
.
\end{eqnarray*}

Let us write \textit{several first} nonlinear PDE's from the Jacobi identity
(\textbf{\ref{full}})%
\begin{eqnarray*}
(\mathcal{W}^{10}+B^{1})\partial _{0}\mathcal{W}^{10}+2\mathcal{W}%
^{11}\partial _{1}\mathcal{W}^{10} &=&2B^{0}\partial _{0}\mathcal{W}%
^{11}+(B^{1}+\mathcal{W}^{10})\partial _{1}\mathcal{W}^{11}+(B^{2}+\mathcal{W%
}^{20})\partial _{2}\mathcal{W}^{11}, \\
&& \\
\partial _{1}\mathcal{W}^{10}\partial _{0}\mathcal{W}^{21} &=&\partial _{1}%
\mathcal{W}^{20}\partial _{0}\mathcal{W}^{11}+\partial _{2}\mathcal{W}%
^{20}\partial _{0}\mathcal{W}^{12}, \\
&& \\
\partial _{0}\mathcal{W}^{10}\partial _{1}\mathcal{W}^{20}+\partial _{1}%
\mathcal{W}^{10}\partial _{1}\mathcal{W}^{21} &=&\partial _{0}\mathcal{W}%
^{20}\partial _{1}\mathcal{W}^{10}+\partial _{1}\mathcal{W}^{20}\partial _{1}%
\mathcal{W}^{11}+\partial _{2}\mathcal{W}^{20}\partial _{1}\mathcal{W}^{12},
\\
&& \\
\partial _{0}\mathcal{W}^{10}\partial _{2}\mathcal{W}^{20}+\partial _{1}%
\mathcal{W}^{10}\partial _{2}\mathcal{W}^{21} &=&\partial _{1}\mathcal{W}%
^{20}\partial _{2}\mathcal{W}^{11}+\partial _{2}\mathcal{W}^{20}\partial _{2}%
\mathcal{W}^{12}, \\
&& \\
\partial _{1}\mathcal{W}^{10}\partial _{3}\mathcal{W}^{21} &=&\partial _{2}%
\mathcal{W}^{20}\partial _{3}\mathcal{W}^{12}.
\end{eqnarray*}%
This system involves just four moments $B^{0},B^{1},B^{2},B^{3}$. It is
enough to find coefficients $\mathcal{W}^{10}$, $\mathcal{W}^{11}$, $%
\mathcal{W}^{20}$. For instance, $\mathcal{W}^{10}$ is a solution of the
Monge--Ampere equation%
\begin{equation*}
(\mathcal{W}^{10})_{00}(\mathcal{W}^{10})_{11}-[(\mathcal{W}%
^{10})_{01}]^{2}+\varphi ^{\prime ^{2}}(B^{0})[(\mathcal{W}^{10})_{1}]^{2}=0%
\text{,}
\end{equation*}%
where $\varphi (B^{0})$ is a some function (determined from the
compatibility conditions $[(\mathcal{W}^{11})_{1}]_{2}=[(\mathcal{W}%
^{11})_{2}]_{1}$, \ $[(\mathcal{W}^{11})_{1}]_{0}=[(\mathcal{W}%
^{11})_{0}]_{1}$, \ $[(\mathcal{W}^{11})_{0}]_{2}=[(\mathcal{W}%
^{11})_{2}]_{0}$, see expressions below).%
\begin{equation*}
\mathcal{W}^{20}=B^{2}(\mathcal{W}^{10})_{1}+G,
\end{equation*}%
where $G(p,q)$ is a solution of the Euler--Darboux--Poisson equation%
\begin{eqnarray*}
G_{pq} &=&\frac{\varphi ^{\prime \prime }(B^{0})}{4\varphi ^{\prime
^{2}}(B^{0})}(G_{p}-G_{q})\text{, \ \ \ \ }\varphi (B^{0})=\frac{1}{2}(p-q)%
\text{,} \\
&& \\
q &=&\ln [(\mathcal{W}^{10})_{1}]-\varphi (B^{0})\text{, \ \ \ \ \ \ }p=\ln
[(\mathcal{W}^{10})_{1}]+\varphi (B^{0}).
\end{eqnarray*}%
Explicit expressions for all other coefficients (depending on higher moments 
$B^{k}$, $k=3,4,5,...$) can be found recursively in \textit{complete}
differentials. For instance, the equation%
\begin{equation*}
(\mathcal{W}^{11})_{2}=\frac{2[(\mathcal{W}^{10})_{1}]^{2}(B^{0}(\mathcal{W}%
^{10})_{00}+(\mathcal{W}^{10}+B^{1})(\mathcal{W}^{10})_{10}+\mathcal{W}^{11}(%
\mathcal{W}^{10})_{11})}{MB^{2}+N},
\end{equation*}%
where%
\begin{eqnarray*}
M &=&[1+(\mathcal{W}^{10})_{1}](\mathcal{W}^{10})_{1}(\mathcal{W}%
^{10})_{11}+2B^{0}[(\mathcal{W}^{10})_{1}(\mathcal{W}^{10})_{110}-(\mathcal{W%
}^{10})_{10}(\mathcal{W}^{10})_{11}] \\
&& \\
&&+(\mathcal{W}^{10}+B^{1})[(\mathcal{W}^{10})_{1}(\mathcal{W}^{10})_{111}-((%
\mathcal{W}^{10})_{11})^{2}], \\
&& \\
N &=&G(\mathcal{W}^{10})_{1}(\mathcal{W}^{10})_{11}+2B^{0}[G_{10}(\mathcal{W}%
^{10})_{1}-G_{1}(\mathcal{W}^{10})_{10}] \\
&& \\
&&+[(\mathcal{W}^{10})_{1}+B^{1}][G_{11}(\mathcal{W}^{10})_{1}-G_{1}(%
\mathcal{W}^{10})_{11}],
\end{eqnarray*}%
can be solved up to a some function of $B^{0}$ and $B^{1}$, which can be
found in complete differentials by a substitution into other derivatives%
\begin{eqnarray*}
(\mathcal{W}^{11})_{1} &=&(\mathcal{W}^{11})_{2}\frac{[(\mathcal{W}%
^{10})_{1}(\mathcal{W}^{10})_{111}-((\mathcal{W}%
^{10})_{11})^{2}]B^{2}+G_{11}(\mathcal{W}^{10})_{1}-G_{1}(\mathcal{W}%
^{10})_{11}}{(\mathcal{W}^{10})_{1}(\mathcal{W}^{10})_{11}}+K, \\
&& \\
(\mathcal{W}^{11})_{0} &=&(\mathcal{W}^{11})_{2}\frac{[(\mathcal{W}%
^{10})_{1}(\mathcal{W}^{10})_{110}-(\mathcal{W}^{10})_{10}(\mathcal{W}%
^{10})_{11}]B^{2}+G_{10}(\mathcal{W}^{10})_{1}-G_{1}(\mathcal{W}^{10})_{10}}{%
(\mathcal{W}^{10})_{1}(\mathcal{W}^{10})_{11}}-L,
\end{eqnarray*}%
where%
\begin{equation*}
K=\frac{(\mathcal{W}^{10})_{0}(\mathcal{W}^{10})_{11}-2(\mathcal{W}%
^{10})_{1}(\mathcal{W}^{10})_{10}}{(\mathcal{W}^{10})_{11}}\text{, \ \ \ \ \
\ \ }L=\frac{(\mathcal{W}^{10})_{1}(\mathcal{W}^{10})_{00}}{(\mathcal{W}%
^{10})_{11}}.
\end{equation*}

Let us write the first two equations from the (\textbf{\ref{mo}}) and the
first equation from (\textbf{\ref{no}})%
\begin{eqnarray*}
B_{t}^{0} &=&(B^{0}\partial _{x}+\partial _{x}B^{0})\frac{\delta \mathbf{%
\bar{H}}_{1}}{\delta B^{0}}+[B^{1}\partial _{x}+\partial _{x}\mathcal{W}%
^{10}(B^{0},B^{1})]\frac{\delta \mathbf{\bar{H}}_{1}}{\delta B^{1}}, \\
&& \\
B_{t}^{1} &=&[\mathcal{W}^{10}(B^{0},B^{1})\partial _{x}+\partial _{x}B^{1}]%
\frac{\delta \mathbf{\bar{H}}_{1}}{\delta B^{0}}+[\mathcal{W}%
^{11}(B^{0},B^{1},B^{2})\partial _{x}+\partial _{x}\mathcal{W}%
^{11}(B^{0},B^{1},B^{2})]\frac{\delta \mathbf{\bar{H}}_{1}}{\delta B^{1}}, \\
&& \\
B_{y}^{0} &=&(B^{0}\partial _{x}+\partial _{x}B^{0})\frac{\delta \mathbf{%
\bar{H}}_{2}}{\delta B^{0}}+[B^{1}\partial _{x}+\partial _{x}\mathcal{W}%
^{10}(B^{0},B^{1})]\frac{\delta \mathbf{\bar{H}}_{2}}{\delta B^{1}}%
+[B^{2}\partial _{x}+\partial _{x}\mathcal{W}^{20}(B^{0},B^{1},B^{2})]\frac{%
\delta \mathbf{\bar{H}}_{2}}{\delta B^{2}}.
\end{eqnarray*}%
These three equations can be written in the conservative form%
\begin{equation*}
\partial _{t}\mathbf{H}_{0}=\partial _{x}F_{00}(\mathbf{H}_{0},\mathbf{H}%
_{1})\text{, \ \ \ \ \ }\partial _{t}\mathbf{H}_{1}=\partial _{x}F_{01}(%
\mathbf{H}_{0},\mathbf{H}_{1},\mathbf{H}_{2})\text{, \ \ \ \ \ \ }\partial
_{y}\mathbf{H}_{0}=\partial _{x}F_{10}(\mathbf{H}_{0},\mathbf{H}_{1},\mathbf{%
H}_{2}).
\end{equation*}%
Let us introduce the potential function $z$, where $\mathbf{H}_{0}=z_{x}$, \ 
$F_{00}(\mathbf{H}_{0},\mathbf{H}_{1})=z_{t}$, \ $F_{10}(\mathbf{H}_{0},%
\mathbf{H}_{1},\mathbf{H}_{2})=z_{y}$. Then one can substitute $\mathbf{H}%
_{1}(z_{x},z_{t})$ and $\mathbf{H}_{2}(z_{x},z_{t},z_{y})$ into the second
equation%
\begin{equation*}
\partial _{t}\mathbf{H}_{1}(z_{x},z_{t})=\partial _{x}F_{01}(\mathbf{H}_{0},%
\mathbf{H}_{1}(z_{x},z_{t}),\mathbf{H}_{2}(z_{x},z_{t},z_{y})).
\end{equation*}%
This 2+1 quasilinear equation of the second order is integrable by the
method of hydrodynamic reductions \textbf{\cite{Fer+Kar}} (or by introducing
pseudopotentials; see again \textbf{\cite{Fer+Kar}} and \textbf{\cite%
{Zakh+multi}}). The Hamiltonian structure of this equation is unknown, while
the Hamiltonian structure of corresponding hydrodynamic chains (\textbf{\ref%
{mo}}) and (\textbf{\ref{no}}) is defined.

\section{$M=1$}

We omit the sub-index in all components $\mathcal{W}_{(1)}^{nk}$ of the
Poisson bracket (see (\textbf{\ref{!}}), $M=1$)%
\begin{equation*}
\{A^{k},A^{n}\}=[\mathcal{W}^{kn}(A^{0},A^{1},...,A^{k+n-1})\partial
_{x}+\partial _{x}\mathcal{W}^{nk}(A^{0},A^{1},...,A^{k+n-1})]\delta
(x-x^{\prime }).
\end{equation*}%
The Jacobi identity (\textbf{\ref{full}}) is%
\begin{eqnarray*}
\underset{m=0}{\overset{n+k-1}{\sum }}(\mathcal{W}^{pm}+\mathcal{W}%
^{mp})\partial _{m}\mathcal{W}^{nk} &=&\underset{m=0}{\overset{n+p-1}{\sum }}%
(\mathcal{W}^{km}+\mathcal{W}^{mk})\partial _{m}\mathcal{W}^{np}, \\
&& \\
\underset{m=0}{\overset{s+p-1}{\sum }}\partial _{m}\mathcal{W}^{sp}\partial
_{n}\mathcal{W}^{km} &=&\underset{m=0}{\overset{p+k-1}{\sum }}\partial _{m}%
\mathcal{W}^{kp}\partial _{n}\mathcal{W}^{sm}.
\end{eqnarray*}%
Thus, the above system of nonlinear PDE's describes a family of the local
Poisson brackets connected with the Hamiltonian hydrodynamic chains.

The existence of commuting hydrodynamic chains%
\begin{eqnarray}
A_{t}^{k} &=&[\mathcal{W}^{kn}(A^{0},A^{1},...,A^{k+n-1})\partial
_{x}+\partial _{x}\mathcal{W}^{nk}(A^{0},A^{1},...,A^{k+n-1})]\frac{\delta 
\mathbf{\bar{H}}_{2}}{\delta A^{n}},  \label{km} \\
&&  \notag \\
A_{y}^{k} &=&[\mathcal{W}^{kn}(A^{0},A^{1},...,A^{k+n-1})\partial
_{x}+\partial _{x}\mathcal{W}^{nk}(A^{0},A^{1},...,A^{k+n-1})]\frac{\delta 
\mathbf{\bar{H}}_{3}}{\delta A^{n}},  \notag
\end{eqnarray}%
where the Hamiltonians are $\mathbf{\bar{H}}_{2}\mathbf{=}\int \mathbf{H}%
_{2}(A^{0},A^{1},A^{2})dx$ and $\mathbf{\bar{H}}_{3}\mathbf{=}\int \mathbf{H}%
_{3}(A^{0},A^{1},A^{2},A^{3})dx$, implies the existence of a hierarchy of
integrable hydrodynamic chains.

The \textit{auxiliary} restrictions (``normalization'')%
\begin{equation*}
\mathcal{W}^{0k}(A^{0},A^{1},...,A^{k-1})\equiv \mathcal{\bar{W}}^{00}\delta
^{0k}\text{, \ \ \ \ \ \ \ \ \ \ \ \ }\mathcal{W}%
^{1k}(A^{0},A^{1},...,A^{k})\equiv A^{k},
\end{equation*}%
where $\mathcal{\bar{W}}^{00}=\limfunc{const}$ (and $\delta ^{ik}$ is the
Kronecker symbol), can be obtained by virtue of an existence of conservation
laws of the Casimir $\mathbf{\bar{H}}_{0}=\int A^{0}dx$ and the momentum $%
\mathbf{\bar{H}}_{1}=\int A^{1}dx$. Then the first three conservation laws
are%
\begin{eqnarray*}
A_{t}^{0} &=&\partial _{x}\left( 2\mathcal{\bar{W}}^{00}\frac{\partial 
\mathbf{H}_{2}}{\partial A^{0}}+A^{0}\frac{\partial \mathbf{H}_{2}}{\partial
A^{1}}+\mathcal{W}^{20}\frac{\partial \mathbf{H}_{2}}{\partial A^{2}}\right) 
\text{,} \\
&& \\
A_{t}^{1} &=&\partial _{x}\left( A^{0}\frac{\partial \mathbf{H}_{2}}{%
\partial A^{0}}+2A^{1}\frac{\partial \mathbf{H}_{2}}{\partial A^{1}}+(%
\mathcal{W}^{21}+A^{2})\frac{\partial \mathbf{H}_{2}}{\partial A^{2}}-%
\mathbf{H}_{2}\right) \text{,} \\
&& \\
\partial _{t}\mathbf{H}_{2} &=&\partial _{x}[\mathcal{\varepsilon }(\frac{%
\partial \mathbf{H}_{2}}{\partial A^{0}})^{2}+A^{0}\frac{\partial \mathbf{H}%
_{2}}{\partial A^{0}}\frac{\partial \mathbf{H}_{2}}{\partial A^{1}}+\mathcal{%
W}^{20}\frac{\partial \mathbf{H}_{2}}{\partial A^{0}}\frac{\partial \mathbf{H%
}_{2}}{\partial A^{2}} \\
&& \\
&&+A^{1}(\frac{\partial \mathbf{H}_{2}}{\partial A^{1}})^{2}+(\mathcal{W}%
^{21}+A^{2})\frac{\partial \mathbf{H}_{2}}{\partial A^{1}}\frac{\partial 
\mathbf{H}_{2}}{\partial A^{2}}+\mathcal{W}^{22}(\frac{\partial \mathbf{H}%
_{2}}{\partial A^{2}})^{2}],
\end{eqnarray*}%
where $\varepsilon \equiv \mathcal{\bar{W}}^{00}$.

Let us write \textit{several first} nonlinear PDE's from the Jacobi identity
(\textbf{\ref{full}})%
\begin{eqnarray*}
\lbrack 2\varepsilon \partial _{0}+A^{0}\partial _{1}+\mathcal{W}%
^{20}\partial _{2}]\mathcal{W}^{21} &=&[A^{0}\partial _{0}+2A^{1}\partial
_{1}]\mathcal{W}^{20} \\
&& \\
\partial _{1}\mathcal{W}^{20}\partial _{0}\mathcal{W}^{31} &=&\partial _{1}%
\mathcal{W}^{30}\partial _{0}\mathcal{W}^{21}+\partial _{2}\mathcal{W}%
^{30}\partial _{0}\mathcal{W}^{22}, \\
&& \\
\partial _{0}\mathcal{W}^{30}\partial _{1}\mathcal{W}^{20}+\partial _{1}%
\mathcal{W}^{30}\partial _{1}\mathcal{W}^{21}+\partial _{2}\mathcal{W}%
^{30}\partial _{1}\mathcal{W}^{22} &=&\partial _{0}\mathcal{W}^{20}\partial
_{1}\mathcal{W}^{30}+\partial _{1}\mathcal{W}^{20}\partial _{1}\mathcal{W}%
^{31}, \\
&& \\
\partial _{0}\mathcal{W}^{20}\partial _{2}\mathcal{W}^{30}+\partial _{1}%
\mathcal{W}^{20}\partial _{2}\mathcal{W}^{31} &=&\partial _{1}\mathcal{W}%
^{30}\partial _{2}\mathcal{W}^{21}+\partial _{2}\mathcal{W}^{30}\partial _{2}%
\mathcal{W}^{22}, \\
&& \\
\partial _{1}\mathcal{W}^{20}\partial _{3}\mathcal{W}^{31} &=&\partial _{2}%
\mathcal{W}^{30}\partial _{3}\mathcal{W}^{22}.
\end{eqnarray*}%
This system involves just four moments $A^{0},A^{1},A^{2},A^{3}$. It is
enough to find coefficients $\mathcal{W}^{20}$, $\mathcal{W}^{21}$, $%
\mathcal{W}^{30}$. For instance, $\mathcal{W}^{20}$ is a solution of the
Monge--Ampere equation%
\begin{equation}
(\mathcal{W}^{20})_{00}(\mathcal{W}^{20})_{11}-[(\mathcal{W}%
^{20})_{01}]^{2}+\varphi ^{\prime ^{2}}(A^{0})[(\mathcal{W}^{20})_{1}]^{2}=0%
\text{,}  \label{ma}
\end{equation}%
where $\varphi (A^{0})$ is a some function (determined from the
compatibility conditions $[(\mathcal{W}^{21})_{1}]_{2}=[(\mathcal{W}%
^{21})_{2}]_{1}$, \ $[(\mathcal{W}^{21})_{1}]_{0}=[(\mathcal{W}%
^{21})_{0}]_{1}$, \ $[(\mathcal{W}^{21})_{0}]_{2}=[(\mathcal{W}%
^{21})_{2}]_{0}$, see these expressions below).%
\begin{equation*}
\mathcal{W}^{30}=A^{2}(\mathcal{W}^{20})_{1}+G,
\end{equation*}%
where $G(p,q)$ is a solution of the Euler--Darboux--Poisson equation%
\begin{eqnarray*}
G_{pq} &=&\frac{\varphi ^{\prime \prime }(A^{0})}{4\varphi ^{\prime
^{2}}(A^{0})}(G_{p}-G_{q})\text{, \ \ \ \ }\varphi (A^{0})=\frac{1}{2}(p-q)%
\text{,} \\
&& \\
q &=&\ln [(\mathcal{W}^{20})_{1}]-\varphi (A^{0})\text{, \ \ \ \ \ \ }p=\ln
[(\mathcal{W}^{20})_{1}]+\varphi (A^{0}).
\end{eqnarray*}%
Explicit expressions for all other coefficients (depending on higher moments 
$A^{k}$, $k=3,4,5,...$) can be found recursively in \textit{complete}
differentials. For instance,%
\begin{equation*}
d\mathcal{W}^{21}=(\mathcal{W}^{21})_{0}dA^{0}+(\mathcal{W}^{21})_{1}dA^{1}+(%
\mathcal{W}^{21})_{2}dA^{2},
\end{equation*}%
where%
\begin{eqnarray*}
(\mathcal{W}^{21})_{2} &=&\frac{2[A^{1}(\mathcal{W}^{20})_{11}+A^{0}(%
\mathcal{W}^{20})_{10}+\varepsilon (\mathcal{W}^{20})_{00}]}{2\varepsilon
\left( \frac{(\mathcal{W}^{20})_{11}A^{2}+G_{1}}{(\mathcal{W}^{20})_{1}}%
\right) _{0}+A^{0}\left( \frac{(\mathcal{W}^{20})_{11}A^{2}+G_{1}}{(\mathcal{%
W}^{20})_{1}}\right) _{1}+\frac{(\mathcal{W}^{20})(\mathcal{W}^{20})_{11}}{(%
\mathcal{W}^{20})_{1}}} \\
&& \\
(\mathcal{W}^{21})_{1} &=&\frac{(\mathcal{W}^{20})_{1}}{(\mathcal{W}%
^{20})_{11}}(\frac{(\mathcal{W}^{20})_{11}A^{2}+G_{1}}{(\mathcal{W}^{20})_{1}%
})_{1}(\mathcal{W}^{21})_{2}+(\mathcal{W}^{20})_{0}-2\frac{(\mathcal{W}%
^{20})_{1}(\mathcal{W}^{20})_{10}}{(\mathcal{W}^{20})_{11}} \\
&& \\
(\mathcal{W}^{21})_{0} &=&\frac{(\mathcal{W}^{20})_{1}}{(\mathcal{W}%
^{20})_{11}}(\frac{(\mathcal{W}^{20})_{11}A^{2}+G_{1}}{(\mathcal{W}^{20})_{1}%
})_{0}(\mathcal{W}^{21})_{2}-\frac{(\mathcal{W}^{20})_{1}(\mathcal{W}%
^{20})_{00}}{(\mathcal{W}^{20})_{11}}.
\end{eqnarray*}

\textbf{Remark}: The coincidence of the coefficients $\mathcal{W}^{20}$ from
this section and $\mathcal{W}^{10}$ from the previous section is easy to
understand if one takes into account that if the Hamiltonian density $%
\mathbf{H}_{2}$ is a function of the moments $A^{1}$ and $A^{2}$ only, then
the corresponding hydrodynamic chain has the Hamiltonian structure
coinciding (by the ``shift'' $A^{k}\rightarrow B^{k-1}$, \ $k=1,2,...$) with
the Hamiltonian structure presented in the previous section.

\textbf{Remark}: The Monge--Ampere equation (\textbf{\ref{ma}}) was derived
in \textbf{\cite{Fer+Dav}}. In this paper solutions of this equation
classify the integrable hydrodynamic chains%
\begin{equation*}
\partial _{t}\mathbf{H}_{0}=\partial _{x}F_{1}(\mathbf{H}_{0},\mathbf{H}_{1})%
\text{,\ \ \ \ }\partial _{t}\mathbf{H}_{1}=\partial _{x}F_{2}(\mathbf{H}%
_{0},\mathbf{H}_{1},\mathbf{H}_{2})\text{, \ \ \ }\partial _{t}\mathbf{H}%
_{2}=\partial _{x}F_{3}(\mathbf{H}_{0},\mathbf{H}_{1},\mathbf{H}_{2},\mathbf{%
H}_{3})\text{, ...}
\end{equation*}%
These hydrodynamic chains \textit{can coincide} with the integrable
hydrodynamic chains (\textbf{\ref{km}}) determined by the Hamiltonian
density $\mathbf{H}_{2}=A^{2}+Q(A^{0},A^{1})$.

All other Hamiltonian hydrodynamic chains (\textbf{\ref{liu}}) can be
investigated in the same way using the Jacobi identity (\textbf{\ref{jac}}).

\section{The Miura type and reciprocal transformations}

Suppose two Hamiltonian hydrodynamic chains%
\begin{equation*}
B_{t}^{k}=\underset{n=0}{\overset{N_{k}}{\sum }}F_{n}^{k}(\mathbf{B}%
)B_{x}^{n}\text{, \ \ \ \ \ \ \ }A_{t}^{k}=\underset{m=0}{\overset{M_{k}}{%
\sum }}G_{n}^{k}(\mathbf{A})A_{x}^{n}\text{,\ \ \ \ \ \ }k=1,2,...,
\end{equation*}%
where $N_{k}$ and $M_{k}$ are some integers, are related by the Miura type
transformations%
\begin{equation*}
A^{k}=A^{k}(B^{0},B^{1},...,B^{k+1})\text{, \ \ \ \ \ \ }k=0,1,2,...
\end{equation*}

\textbf{Theorem}: \textit{The corresponding Poisson brackets} (\textit{see} (%
\textbf{\ref{has}}) \textit{and} (\textbf{\ref{!}}))%
\begin{eqnarray}
\{B^{k},B^{n}\} &=&[G_{(L)}^{kn}(B^{0},B^{1},...,B^{k+n-L})\partial
_{x}+\Gamma _{(L)m}^{kn}(\mathbf{B})B_{x}^{m}]\delta (x-x^{\prime }),
\label{fi} \\
&&  \notag \\
\{A^{k},A^{n}\} &=&[G_{(M)}^{kn}(A^{0},A^{1},...,A^{k+n-M})\partial
_{x}+\Gamma _{(M)m}^{kn}(\mathbf{A})A_{x}^{m}]\delta (x-x^{\prime })
\label{ch}
\end{eqnarray}%
\textit{are related by the above Miura type transformations if} $L=M+1$.

\textbf{Proof}: It is easy to check by taking into account the \textit{%
highest order} dependence on the moments $B^{n}$ only.%
\begin{equation*}
\{A^{k},A^{n}\}=\underset{i=0}{\overset{k+1}{\sum }}\underset{j=0}{\overset{%
n+1}{\sum }}\frac{\partial A^{k}}{\partial B^{i}}\{B^{i},B^{j}\}\frac{%
\partial A^{n}}{\partial B^{j}}\delta (x-x^{\prime })\sim \frac{\partial
A^{k}}{\partial B^{k+1}}\{B^{k+1},B^{n+1}\}\frac{\partial A^{n}}{\partial
B^{n+1}}\delta (x-x^{\prime }).
\end{equation*}%
Thus, $G_{(M)}^{kn}(A^{0},A^{1},...,A^{k+n-M})\sim
G_{(L)}^{k+1,n+1}(B^{0},B^{1},...,B^{k+n+2-L})$. So, indeed, $L=M+1$.

\textbf{Remark}: Suppose that both coordinate systems are the Liouville
coordinates. Then, one obtains the linear system describing such Miura type
transformations%
\begin{equation*}
\partial _{m}\left[ \frac{\partial A^{k}}{\partial B^{i}}[(\mathcal{W}%
_{(L)}^{ij}+\mathcal{W}_{(L)}^{ji})\frac{\partial ^{2}A^{n}}{\partial
B^{j}\partial B^{s}}+\mathcal{W}_{(L)s}^{ji}\frac{\partial A^{n}}{\partial
B^{j}}]\right] =\partial _{s}\left[ \frac{\partial A^{k}}{\partial B^{i}}[(%
\mathcal{W}_{(L)}^{ij}+\mathcal{W}_{(L)}^{ji})\frac{\partial ^{2}A^{n}}{%
\partial B^{j}\partial B^{m}}+\mathcal{W}_{(L)m}^{ji}\frac{\partial A^{n}}{%
\partial B^{j}}]\right] .
\end{equation*}%
If the first Hamiltonian structure written in the Liouville coordinates (%
\textbf{\ref{fi}}) and the Miura type transformations (\textbf{\ref{!}}) are
given, then the second Hamiltonian structure (\textbf{\ref{ch}}) can be
reconstructed in complete differentials, where%
\begin{equation*}
\frac{\partial \mathcal{W}_{(M)}^{nk}(\mathbf{A})}{\partial B^{s}}=\frac{%
\partial A^{k}}{\partial B^{i}}[(\mathcal{W}_{(L)}^{ij}+\mathcal{W}%
_{(L)}^{ji})\frac{\partial ^{2}A^{n}}{\partial B^{j}\partial B^{s}}+\mathcal{%
W}_{(L)s}^{ji}\frac{\partial A^{n}}{\partial B^{j}}].
\end{equation*}

\textbf{Example}: The Benney hydrodynamic chain (see (\textbf{\ref{kmb}}), 
\textbf{\cite{Benney}} and \textbf{\cite{KM}})%
\begin{equation*}
A_{t}^{k}=A_{x}^{k+1}+kA^{k-1}A_{x}^{0}=[kA^{k+n-1}\partial _{x}+n\partial
_{x}A^{k+n-1}]\frac{\partial \mathbf{H}_{2}}{\partial A^{n}}\text{, \ \ \ \ }%
k=0,1,2,...,
\end{equation*}%
where the Hamiltonian density $\mathbf{H}_{2}=A^{2}+(A^{0})^{2}$, is
connected with the modified Benney hydrodynamic chain (see \textbf{\cite%
{Teshuk}}, the particular case of the Kupershmidt hydrodynamic chains 
\textbf{\cite{Kuper}})%
\begin{equation*}
B_{t}^{k}=B_{x}^{k+1}+B^{0}B_{x}^{k}+(k+2)B^{k}B_{x}^{0}=[(k+1)B^{k+n}%
\partial _{x}+(n+1)\partial _{x}B^{k+n}]\frac{\partial \mathbf{H}_{0}}{%
\partial B^{n}},
\end{equation*}%
where the Hamiltonian density $\mathbf{H}_{0}=B^{1}+(B^{0})^{2}$, by the
Miura type transformations (see \textbf{\cite{Maks+wdvv}})%
\begin{equation*}
A^{0}=B^{1}+(B^{0})^{2}\text{, \ }A^{1}=B^{2}+3B^{0}B^{1}+2(B^{0})^{3},...
\end{equation*}%
In this case $L=0$ and $M=-1$. Thus, the Benney moment chain has the second
local Hamiltonian structure determined by the Poisson bracket (\textbf{\ref%
{ch}}), whose coefficients can be found by using the above Miura type
transformations.

\textbf{Remark}: In the theory of dispersive integrable systems the Miura
transformation is a tool to reduce the Hamiltonian structure to canonical
form ``$d/dx$'' (the infinitely many component analogue of the Darboux
theorem, see \textbf{\cite{Maks+kdv}}). In the theory of the Poisson
brackets associated with hydrodynamic chains an application of the Miura
type transformation reduces the dependence on a number of the moments $A^{k}$%
. It means that the following diagram exists (see (\textbf{\ref{has}) and }(%
\textbf{\ref{!}}))%
\begin{eqnarray*}
\{A^{k},A^{n}\} &=&[G_{(-M)}^{kn}(A^{0},...,A^{k+n+M})+...]\delta
(x-x^{\prime }), \\
&\downarrow & \\
\{B^{k},B^{n}\} &=&[G_{(1-M)}^{kn}(B^{0},...,B^{k+n+M-1})+...]\delta
(x-x^{\prime }), \\
&\downarrow & \\
\{C^{k},C^{n}\} &=&[G_{(2-M)}^{kn}(C^{0},...,C^{k+n+M-2})+...]\delta
(x-x^{\prime }), \\
&\downarrow & \\
&&.....
\end{eqnarray*}

\textbf{Conjecture}: Possibly, the number of such Miura type transformations
is \textit{infinite}, and any local Hamiltonian structure with the index $-M$
can be reduced to local Hamiltonian structure with an arbitrary index $N$,
where the triangular block $\mathcal{\bar{W}}_{(N)}^{kn}=\limfunc{const}$
increases (see (\textbf{\ref{2}}), if $N$ increases) in complete accordance
with the Darboux theorem.

Let us apply the reciprocal transformation%
\begin{equation*}
dz=\mathbf{F}(A^{0},A^{1},...)dx+\mathbf{G}(A^{0},A^{1},...)dt\text{, \ \ \
\ \ \ }dy=dt
\end{equation*}%
to the hydrodynamic chain (\textbf{\ref{j}})%
\begin{equation*}
A_{t}^{k}=\underset{n=0}{\overset{M+1}{\sum }}V_{n}^{k}(\mathbf{A})A_{x}^{n}%
\text{, \ \ }k=0,1,...,M-1\text{, \ \ \ }A_{t}^{k}=\underset{n=0}{\overset{%
k+1}{\sum }}V_{n}^{k}(\mathbf{A})A_{x}^{n}\text{, \ \ }k=M,M+1,...
\end{equation*}%
If the conservation law density $\mathbf{F}(A^{0},A^{1},...)$ is a linear
function (with special choice of constants, see \textbf{\cite{Fer+trans}})
of $M$ Casimirs $A^{k}$ and the momentum $A^{M}$, then new hydrodynamic chain%
\begin{equation*}
A_{y}^{k}=\underset{n=0}{\overset{M+1}{\sum }}(\mathbf{F}V_{n}^{k}(\mathbf{A}%
)-\mathbf{G}\delta _{n}^{k})A_{z}^{n}\text{,\ \ }k=0,1,...,M-1\text{, \ \ \ }%
A_{y}^{k}=\underset{n=0}{\overset{k+1}{\sum }}(\mathbf{F}V_{n}^{k}(\mathbf{A}%
)-\mathbf{G}\delta _{n}^{k})A_{z}^{n}\text{,\ \ }k=M,M+1,...
\end{equation*}%
preserves the \textit{local} Hamiltonian structure (see (\textbf{\ref{has}})
and\textbf{\ }(\textbf{\ref{!}})), i.e.%
\begin{eqnarray*}
\{A^{k}(x),A^{n}(x^{\prime })\}
&=&[G_{(M)}^{kn}(A^{0},...,A^{k+n-M})+...]\delta (x-x^{\prime }), \\
&\downarrow & \\
\{A^{k}(z),A^{n}(z^{\prime })\} &=&[\tilde{G}%
_{(M)}^{kn}(A^{0},...,A^{k+n-M})+...]\delta (z-z^{\prime }).
\end{eqnarray*}%
Possibly, more general (generalized) reciprocal transformations (see \textbf{%
\cite{Fer+trans}}\ and \textbf{\cite{Fer+Max}}) preserving local Hamiltonian
structures for integrable hydrodynamic chains can be found.

\section{Nonlocal Hamiltonian structures}

The nonlocal Poisson brackets (cf. (\textbf{\ref{ha}}))%
\begin{equation}
\{U^{i},U^{j}\}=[G^{ij}(\mathbf{U})\partial _{x}+\Gamma _{k}^{ij}(\mathbf{U}%
)U_{x}^{k}+\varepsilon U_{x}^{i}\partial _{x}^{-1}U_{x}^{j}]\delta
(x-x^{\prime })\text{, \ \ \ \ \ \ }i,j,k=1,2,3,...  \label{n}
\end{equation}%
for $N$ component case were completely investigated by E.V. Ferapontov and
O.I. Mokhov in \textbf{\cite{Fer+Mokh}} for the non-degenerate matrix $%
G^{ij} $; its degenerate case was considered by O.I. Mokhov in \textbf{\cite%
{Mokh6}}).The Jacobi identity yields the set of restrictions%
\begin{equation*}
G^{ij}=G^{ji}\text{, \ \ \ \ \ \ \ \ \ \ \ \ \ }\partial _{k}G^{ij}=\Gamma
_{k}^{ij}+\Gamma _{k}^{ji}\text{, \ \ \ \ \ \ \ \ \ \ \ \ \ }G^{ik}\Gamma
_{k}^{jn}=G^{jk}\Gamma _{k}^{in},
\end{equation*}%
\begin{equation*}
\varepsilon (G^{im}\delta _{k}^{j}-G^{ij}\delta _{k}^{m})=\Gamma
_{n}^{ij}\Gamma _{k}^{nm}-\Gamma _{n}^{im}\Gamma _{k}^{nj}+G^{in}(\partial
_{n}\Gamma _{k}^{mj}-\partial _{k}\Gamma _{n}^{mj}),
\end{equation*}%
\begin{eqnarray*}
&&-\varepsilon \lbrack (\Gamma _{k}^{ij}-\Gamma _{k}^{ji})\delta
_{p}^{m}+(\Gamma _{k}^{mi}-\Gamma _{k}^{im})\delta _{p}^{j}+(\Gamma
_{k}^{jm}-\Gamma _{k}^{mj})\delta _{p}^{i} \\
&& \\
&&+(\Gamma _{p}^{ij}-\Gamma _{p}^{ji})\delta _{k}^{m}+(\Gamma
_{p}^{mi}-\Gamma _{p}^{im})\delta _{k}^{j}+(\Gamma _{p}^{jm}-\Gamma
_{p}^{mj})\delta _{k}^{i}]
\end{eqnarray*}%
\begin{equation*}
=
\end{equation*}%
\begin{eqnarray*}
&&(\partial _{n}\Gamma _{k}^{ij}-\partial _{k}\Gamma _{n}^{ij})\Gamma
_{p}^{nm}+(\partial _{n}\Gamma _{k}^{mi}-\partial _{k}\Gamma
_{n}^{mi})\Gamma _{p}^{nj}+(\partial _{n}\Gamma _{k}^{jm}-\partial
_{k}\Gamma _{n}^{jm})\Gamma _{p}^{ni} \\
&& \\
&&+(\partial _{n}\Gamma _{p}^{ij}-\partial _{p}\Gamma _{n}^{ij})\Gamma
_{k}^{nm}+(\partial _{n}\Gamma _{p}^{mi}-\partial _{p}\Gamma
_{n}^{mi})\Gamma _{k}^{nj}+(\partial _{n}\Gamma _{p}^{jm}-\partial
_{p}\Gamma _{n}^{jm})\Gamma _{k}^{ni},
\end{eqnarray*}%
which in the Liouville coordinates $A^{i}=A^{i}(\mathbf{U})$ determined by
the conditions (see \textbf{\cite{Malt+Nov}}, \textbf{\cite{Mokh6}})%
\begin{equation*}
G^{ij}=\mathcal{W}^{ij}+\mathcal{W}^{ji}-\varepsilon A^{i}A^{j}\text{, \ \ \
\ \ \ \ \ \ }\Gamma _{k}^{ij}=\partial _{k}\mathcal{W}^{ji}-\varepsilon
\delta _{k}^{i}A^{j},
\end{equation*}%
simplify to (cf. (\textbf{\ref{jac}}))%
\begin{eqnarray}
(\mathcal{W}^{ik}+\mathcal{W}^{ki}-\varepsilon A^{i}A^{k})\partial _{k}%
\mathcal{W}^{nj} &=&(\mathcal{W}^{jk}+\mathcal{W}^{kj}-\varepsilon
A^{j}A^{k})\partial _{k}\mathcal{W}^{ni},  \notag \\
&&  \label{fu} \\
\partial _{n}\mathcal{W}^{ij}\partial _{m}\mathcal{W}^{kn} &=&\partial _{n}%
\mathcal{W}^{kj}\partial _{m}\mathcal{W}^{in}.  \notag
\end{eqnarray}%
Thus, the nonlocal Poisson bracket in the Liouville coordinates has the form
(the numeration of the moments $A^{k}$ runs from $0$ up to infinity)%
\begin{equation*}
\{A^{i},A^{j}\}=[\mathcal{W}^{ij}\partial _{x}+\partial _{x}\mathcal{W}%
^{ji}-\varepsilon A^{j}\partial _{x}A^{i}+\varepsilon A_{x}^{i}\partial
_{x}^{-1}A_{x}^{j}]\delta (x-x^{\prime })\text{, \ \ \ \ \ \ }i,j=0,1,...
\end{equation*}%
Similar classification of these nonlocal Poisson brackets (see (\textbf{\ref%
{fu}})) will be made elsewhere.

The first example of such a nonlocal Poisson bracket (\textbf{\ref{n}}) is
(see \textbf{\cite{Maks+Kuper}})%
\begin{eqnarray*}
\{U^{k},U^{n}\} &=&[\beta (\beta k+\beta +1)U^{k+n+1}\partial _{x}+\beta
(\beta n+\beta +1)\partial _{x}U^{k+n+1} \\
&&
\end{eqnarray*}%
\begin{equation*}
+(\beta k+\beta +2)(\beta n+\beta +2)U^{k}U^{n}\partial _{x}+(\beta k+\beta
+2)(\beta n+\beta +1)U^{k}(U^{n})_{x}
\end{equation*}%
\begin{eqnarray*}
&& \\
&&+(\beta n+\beta +2)U^{n}(U^{k})_{x}-(U^{k})_{x}\partial
_{x}^{-1}(U^{n})_{x}]\delta (x-x^{\prime }).
\end{eqnarray*}

\textbf{Remark}: A more general Poisson brackets have been introduced by
E.V. Ferapontov (see \textbf{\cite{Fer+trans}}) for $N$ component
non-degenerate case. A similar extension as in the above case one can
develop for the infinite component case. Then the Poisson bracket in the
Liouville coordinates is (see \textbf{\cite{Mokh+non}})%
\begin{eqnarray*}
\{U^{k},U^{n}\} &=&[(\Phi ^{kn}+\Phi ^{nk}-\varepsilon
_{pq}V^{(p)k}V^{(q)n})\partial _{x}+(\partial _{m}\Phi ^{nk}-\varepsilon
_{pq}V^{(q)n}\partial _{m}V^{(p)k})A_{x}^{m} \\
&& \\
&&+\varepsilon _{pq}\partial _{m}V^{(p)k}A_{x}^{m}\partial _{x}^{-1}\partial
_{s}V^{(q)n}A_{x}^{s}]\delta (x-x^{\prime }),
\end{eqnarray*}%
where $\Phi ^{kn}$ and $V^{(p)k}$ are functions of the moments $A^{m}$, the
constant $L$x$L$ matrix $\varepsilon _{pq}$ is symmetric and non-degenerated
($L$ is ``co-dimension'' of the pseudo-Riemannian space).

\section{Conclusion and outlook}

The \textbf{main claim} of this paper is that the Poisson brackets%
\begin{equation*}
\{A^{k},A^{n}\}=[\mathcal{W}_{(M)}^{kn}(A^{0},...,A^{k+n-M})\partial
_{x}+\partial _{x}\mathcal{W}_{(M)}^{nk}(A^{0},...,A^{k+n-M})]\delta
(x-x^{\prime })\text{, \ \ }k,n=0,1,2,...
\end{equation*}%
can be classified completely by virtue of the Jacobi identity (\textbf{\ref%
{jac}}), which is the over-determined system of infinitely many nonlinear
PDE's. The \textbf{main observation} is that the solution of the several
first nonlinear PDE's allows one to compute all other coefficients
recursively in complete differentials.

Nevertheless, the problem of complete description of these Poisson brackets
is solvable, because all integrable hydrodynamic chains (\textbf{\ref{chain}}%
) are already found (see \textbf{\cite{Fer+Dav}}). Thus, one should just
identify this classification and hydrodynamic chains generated by the
Poisson brackets described above.

\textbf{Conjecture}: \textit{Moreover, we believe that }\textbf{\textit{any
integrable hydrodynamic chain}}%
\begin{equation*}
A_{t}^{k}=\underset{n=0}{\overset{N_{k}}{\sum }}V_{n}^{k}(\mathbf{A}%
)A_{x}^{n}\text{, \ \ \ \ \ }k=0,1,2,...,
\end{equation*}%
\textit{where set of functions} $V_{n}^{k}(\mathbf{A})$ \textit{depends on
the moments} $A^{m}$ ($m=0,1,...,M_{k}$; $M_{k}$ \textit{and} $N_{k}$ 
\textit{are arbitrary integers), possesses \textbf{infinitely many local
Hamiltonian structures} }(\textbf{\ref{liu}}).

Let us consider a bi-Hamiltonian structure (see, for instance, (\textbf{\ref%
{kmb}}) and below) determined by the couple of $M-$brackets%
\begin{equation*}
A_{t}^{k}=[\mathcal{W}_{(M)}^{kn}\partial _{x}+\partial _{x}\mathcal{W}%
_{(M)}^{nk}]\frac{\delta \mathbf{H}_{M+1}}{\delta A^{n}}=[G_{(M+1)}^{kn}+%
\Gamma _{(M+1)s}^{kn}A_{x}^{s}]\frac{\delta \mathbf{H}_{M+2}}{\delta A^{n}}
\end{equation*}%
or, for instance,%
\begin{equation*}
A_{t}^{k}=[\mathcal{W}_{(M)}^{kn}\partial _{x}+\partial _{x}\mathcal{W}%
_{(M)}^{nk}]\frac{\delta \mathbf{H}_{M+1}}{\delta A^{n}}=[\tilde{G}%
_{(M-1)}^{kn}+\tilde{\Gamma}_{(M-1)s}^{kn}A_{x}^{s}]\frac{\delta \mathbf{H}%
_{M}}{\delta A^{n}}.
\end{equation*}%
Suppose, that all $M-$brackets are classified already. If any pair of such
Poisson brackets is compatible (see, for instance, \textbf{\cite{Magri}}),
then this pair automatically creates an integrable hydrodynamic chain
together with recursion (hereditary) operator.

\section*{Acknowledgement}


I thank Gennady El, Eugeni Ferapontov, Andrey Maltsev and Sergey Tsarev for
their suggestions significantly improving my results.

I am grateful to the Institute of Mathematics in Taipei (Taiwan) where some
part of this work has been done, and especially to Jen-Hsu Chang, Jyh-Hao
Lee, Ming-Hien Tu and Derchyi Wu for fruitful discussions.

\end{document}